\documentclass[twocolumn]{webofc}
\usepackage[varg]{txfonts}   
\newcommand{\PPCF}[0]{Plasma Phys. Control. Fusion}
\newcommand{\PoP}[0]{Phys. Plasmas}
\newcommand{\NF}[0]{Nucl. Fusion}

\newcommand{\CPC}[0]{Comp. Phys. Comm.}

\newcommand{\PRL}[0]{Phys. Rev. Lett.}
\usepackage{hyperref}
\begin{document}
\title{The deteriorating effect of plasma density fluctuations on microwave beam quality}

\author{\firstname{Alf} \lastname{Köhn}\inst{1,2}\fnsep\thanks{\email{koehn@igvp.uni-stuttgart.de}} \and
        \firstname{Max E.} \lastname{Austin}\inst{3} \and
        \firstname{Michael W.} \lastname{Brookman}\inst{3} \and
        \firstname{Kenneth W.} \lastname{Gentle}\inst{3} \and        
        \firstname{Lorenzo} \lastname{Guidi}\inst{2} \and
        \firstname{Eberhard} \lastname{Holzhauer}\inst{1} \and
        \firstname{Rob J.} \lastname{La Haye}\inst{4} \and
        \firstname{Jarrod B.} \lastname{Leddy}\inst{5} \and
        \firstname{Omar} \lastname{Maj}\inst{2} \and
        \firstname{Craig C.} \lastname{Petty}\inst{4} \and
        \firstname{Emanuele} \lastname{Poli}\inst{2} \and
        \firstname{Terry L.} \lastname{Rhodes}\inst{6} \and
        \firstname{Antti} \lastname{Snicker}\inst{2,7} \and
        \firstname{Matthew B.} \lastname{Thomas}\inst{5} \and
        \firstname{Roddy G. L.} \lastname{Vann}\inst{5} \and
        \firstname{Hannes} \lastname{Weber}\inst{2}
}

\institute{Institute of Interfacial Process Engineering and Plasma Technology, University of Stuttgart, Stuttgart, Germany
\and
           Max Planck Institute for Plasma Physics, Garching, Germany
\and
			Institute for Fusion Studies, University of Texas at Austin, TX, USA
\and
			General Atomics, PO Box 85608, San Diego, CA, USA
\and
           York Plasma Insitute, Department of Physics, University of York, York, U.K.
\and
			Department of Physics and Astronomy, University of California Los Angeles, Los Angeles, CA, USA
\and
			Department of Applied Physics, Aalto University, Aalto, Finland
          }

\abstract{%
Turbulent plasma edge density fluctuations can broaden a traversing microwave beam degrading its quality. This can be a problem for scenarios relying on a high spatial localization of the deposition of injected microwave power, like controlling MHD instabilities. Here we present numerical estimations of the scattering of a microwave beam by density fluctuations over a large parameter range, including extrapolations to ITER. Two codes are used, the full-wave code IPF-FDMC and the wave kinetic equation solver WKBeam. A successful comparison between beam broadening obtained from DIII-D experiments and corresponding full-wave simulations is shown.
}
\maketitle
%
\section{Introduction}\label{s:intro}
Electromagnetic waves in the microwave frequency range have become an indispensable tool for fusion experiments based on the magnetic confinement concepts of tokamak and stellarator. \textit{Electron cyclotron resonance heating} (ECRH) and \textit{current drive} (ECCD), see e.g.~Refs.~\cite{Bornatici:1983,Erckmann:1994}, allow to transfer power in the MW-regime~\cite{Thumm:2003} to the plasma. On the other hand, active and passive microwave diagnostics occupying only very little space on the inner wall are routinely used in today's experiment~\cite{Hartfuss:2013} and will become even more important in future reactors~\cite{Volpe.2017}.

Heating and diagnostics suffer, however, both from plasma density fluctuations existing at the plasma boundary. These fluctuations can reach levels of up to $100\,\%$~\cite{Zweben.2007} and potentially spoil heating efficiencies and result in ambiguous diagnostics results. This is in particular a problem for the stabilization of so-called \textit{neo-classical tearing modes} (NTMs), a magneto-hydrodynamic instability arising from small perturbations in the plasma current profile~\cite{LaHaye:2006,Zohm:2007}. If not taken care of, they can result in disruptions which are to be avoided at all costs in large-scale tokamaks like ITER. One way to stabilize the NTMs consists in localized current drive in order to restore the original current profile. ECCD has been successfully used to provide this current~\cite{Kasparek:2016}. Using numerical tools to evaluate the consequences of edge density turbulence on the quality of a microwave beam injected for NTM stabilization is the topic of this paper.

The interaction of electromagnetic waves with plasma density fluctuations is an important topic since the very beginning of plasma physics: in the 1930, radio waves emitted from ground were found to be strongly disturbed after reflection at the ionosphere~\cite{Booker:1950}. Likewise, radio waves emitted by distant rotating neutron stars experience phase disturbances when passing the terrestrial ionosphere which makes detection on Earth complicated. In the 1950s, this problem, which is still an issue for today's satellite communication~\cite{Andrews:1995}, was tackled by describing the ionosphere as a layer of thin phase screens~\cite{Hewish:1951}. This approach relies on small perturbation levels, and locally strong density irregularities provide still a challenge in modeling wave propagation across the ionosphere~\cite{Strangeways:2014}.


In the 1980s high-power microwave sources became available and the interaction of injected microwaves and plasma edge density fluctuations in fusion-relevant scenarios was studied with geometrical-optics tools~\cite{Ott:1980,Hansen:1988}. Due to the potentially negative consequences for NTM stabilization in ITER, this topic has gained significant traction since 2009~\cite{Tsironis:2009,Peysson:2011,Ram:2016,Ioannidis:2017,Snicker:2018,Chellai:2018}. 

\section{Numerical tools}\label{s:numerics}

\subsection{The full-wave code IPF-FDMC}\label{s:fullwave}
The full-wave code IPF-FDMC is based on the finite-difference time-domain scheme. It solves Maxwell's equation and the fluid equation of motion of the electrons on a 2D Cartesian grid. The evolution equations for the wave magnetic field $\mathbf{B}$, the wave electric field $\mathbf{E}$, and the current density $\mathbf{J}$ of the wave read:
\begin{eqnarray}
	\frac{\partial}{\partial t} \mathbf{B} &=& -\nabla \times \mathbf{E}\\
	\frac{\partial}{\partial t} \mathbf{E} &=& c^2\nabla \times \mathbf{B} - \mathbf{J}/\epsilon_0\\
	\frac{\partial}{\partial t} \mathbf{J} &=& \epsilon_0 \omega_{pe}^2 \mathbf{E}
												- \omega_{ce} \mathbf{J}\times\mathbf{\hat{B}_0}
												- \nu_e \mathbf{J},
\end{eqnarray}
with $c$ the speed of light, $\epsilon_0$ the vacuum permittivity, $\omega_{pe}=\sqrt{n_ee^2/(\epsilon_0m_e)}$ the electron plasma frequency, $\omega_{ce}=|e|B_0/m_e$ the electron cyclotron frequency, $\mathbf{\hat{B}_0}$ the unit vector into the direction of the background magnetic field, and $\nu_e$ an electron collision frequency (as a dissipation mechanism). More details about the code can be found in Refs.~\cite{Koehn:2010a,Koehn:2018a}.

\subsection{The WKBeam code}\label{s:wkbeam}
The WKBeam code solves the wave kinetic equation in the presence of random density fluctuations embedded in a slowly varying background plasma density. The fluctuations are included by applying a scattering operator whose derivation is based on the \textit{Born approximation}~\cite{Born:1999}. The wave kinetic equation to be solved can be cast into a form such that the scattering operator depends only on the \textit{correlation function} of the random density field and thus remains valid even for short-scale fluctuations~\cite{McDonald:1991}. Details about the WKBeam code and the derivation of the underlying equations can be found in Refs.~\cite{Snicker:2018,Weber:2015,Koehn:2018a}

Since the derivation of the scattering operator relies on the Born approximation, it is expected to become invalid for large fluctuation levels. More precisely, 
\begin{equation}\label{e:wkbeam_born}
	\left[\langle \left( \delta n_e/n_{e,0} \right)^2 \rangle\right]^{1/2} \omega_{pe}^2/\omega_0^2 \ll 1
\end{equation}
needs to be fulfilled for the Born approximation to hold. Therefore the combination of fluctuation level and background density needs to be taken into account. The validation of the Born approximation in WKBeam is investigated in detail in Ref.~\cite{Koehn:2018a}.

\subsection{The simulation set-up}
To benchmark WKBeam with the full-wave code IPF-FDMC, a 2D computational domain is chosen resembling part of a poloidal cross section in a toroidal magnetic confinement device. The background plasma density profile is linearly increasing as
\begin{equation}\label{e:density_1D}
	n_{e,0}(x) = \frac{n_{e,\mathrm{max}}}{x_1 - x_2} \left( x_1 - x\right),
\end{equation}
with $n_{e,\mathrm{max}} = 0.65\cdot n_{e,\mathrm{cut}}$ (where $n_{e,\mathrm{cut}}$ is the O-mode cut-off density of the injected microwave), $x$ the radial coordinate, $x_1=2.45\,\mathrm{m}$ the position where the density profile starts to rise until a position of $x_2=2.30\,\mathrm{m}$. A layer of fluctuations is added to the background profile, where the envelope of the fluctuation amplitude $F(x)$ is described by a Gaussian. The full electron plasma density profiles then reads
\begin{equation}
	n_e(x,z) = n_{e,0}(x) \left( 1 + F(x) \delta n_e(x,z) \right),
\end{equation}
with $z$ the vertical coordinate and $\delta n_e(x,z)$ the random field which is generated by a truncated sum of Fourier-like modes:
\begin{equation}
	\delta n_e(x,z) = \sum_i^{M_i} \sum_j^{M_j} A_{i,j} \cos\left[ k_{x,i}x + k_{z,j}z + \varphi_{i,j} \right],
\end{equation} 
with $A_{i,j}$ the amplitudes of the modes and $\varphi_{i,j}$ independent random phases. There is no dependence on time as the turbulent density fluctuations appear to be frozen in the time frame of the microwave. 

Figure~\ref{f:movie_snapshot} shows a contour plot of the resulting plasma density. Note that this is an actual sample used in the full-wave simulations as input. To properly simulate the effect of fluctuations in the full-wave code, an ensemble-average is required with the ensemble consisting of a series of density profiles each being a 
sample of the same random density field. We use synthetic turbulence to ensure that the statistics of the random field are the same as those assumed in WKBeam. 

A Gaussian beam in O-mode polarization is injected with its focal point located inside of the grid (as can be seen in Fig.~\ref{f:movie_snapshot}). The simulation parameters are listed in Table~\ref{t:simulation_AUG}. They correspond to ASDEX Upgrade parameters scaled-down by approximately a factor of 3 (the reduction in frequency and hence increase in wavelength allows for coarser numerical grids to be used, strongly reducing the required computational resources).
\begin{table}
\centering
\caption{Simulation parameters for the benchmark together with typical ASDEX-Upgrade values}
\label{t:simulation_AUG}       
\begin{tabular}{rll}
\hline
	 			& benchmark & AUG  \\\hline
	$f$			& $50\,\mathrm{GHz}$ 	& $140\,\mathrm{GHz}$ \\
	$|B_0|$		& $1\,\mathrm{T}$		& $2.5-3\,\mathrm{T}$ \\
	$L_\perp$	& $5\,\mathrm{mm}$		& $2-10\,\mathrm{mm}$ \\
	$w_0$		& $1\,\mathrm{cm}$		& $3\,\mathrm{cm}$ \\\hline
\end{tabular}
\end{table}

\subsection{Data analysis}
The efficiency of the NTM stabilization depends on the good localization of the current driven by the injected microwaves, as explained in the introduction. If the power deposition is too broad, the resulting current filament will be too broad to restore the original plasma current profile. Assuming as a first order approximation that the power deposition width corresponds to the beam width of the microwave at the location of absorption (for a detailed discussion about the correlation between microwave beam width and deposition profile width, see Ref.~\cite{Poli:2015}), the additional broadening of the beam, as compared to the case without turbulence, is chosen as the quantity of interest to be compared between the two codes. 

As a first step, both codes were compared for beams propagating in vacuum, which yielded excellent agreement. Next, the plasma profiles as described by Eq.~(\ref{e:density_1D}) were considered, i.e. without turbulent density fluctuations (again, excellent agreement is found~\cite{Koehn:2018a}). The resulting beam widths are used as reference positions to which the beam widths obtained from scenarios with fluctuations included are compared. In detail, Gaussians are fitted to the ensemble-averaged beam cross sections with the beam width being one fit parameter. The obtained values are then normalized to the cases without fluctuations.

\section{Simulation results}\label{s:simulation_results}

\subsection{Illustration of beam broadening}\label{s:BenK_illustration}
\begin{figure}[ht]
\centering
\includegraphics[width=0.45\textwidth]{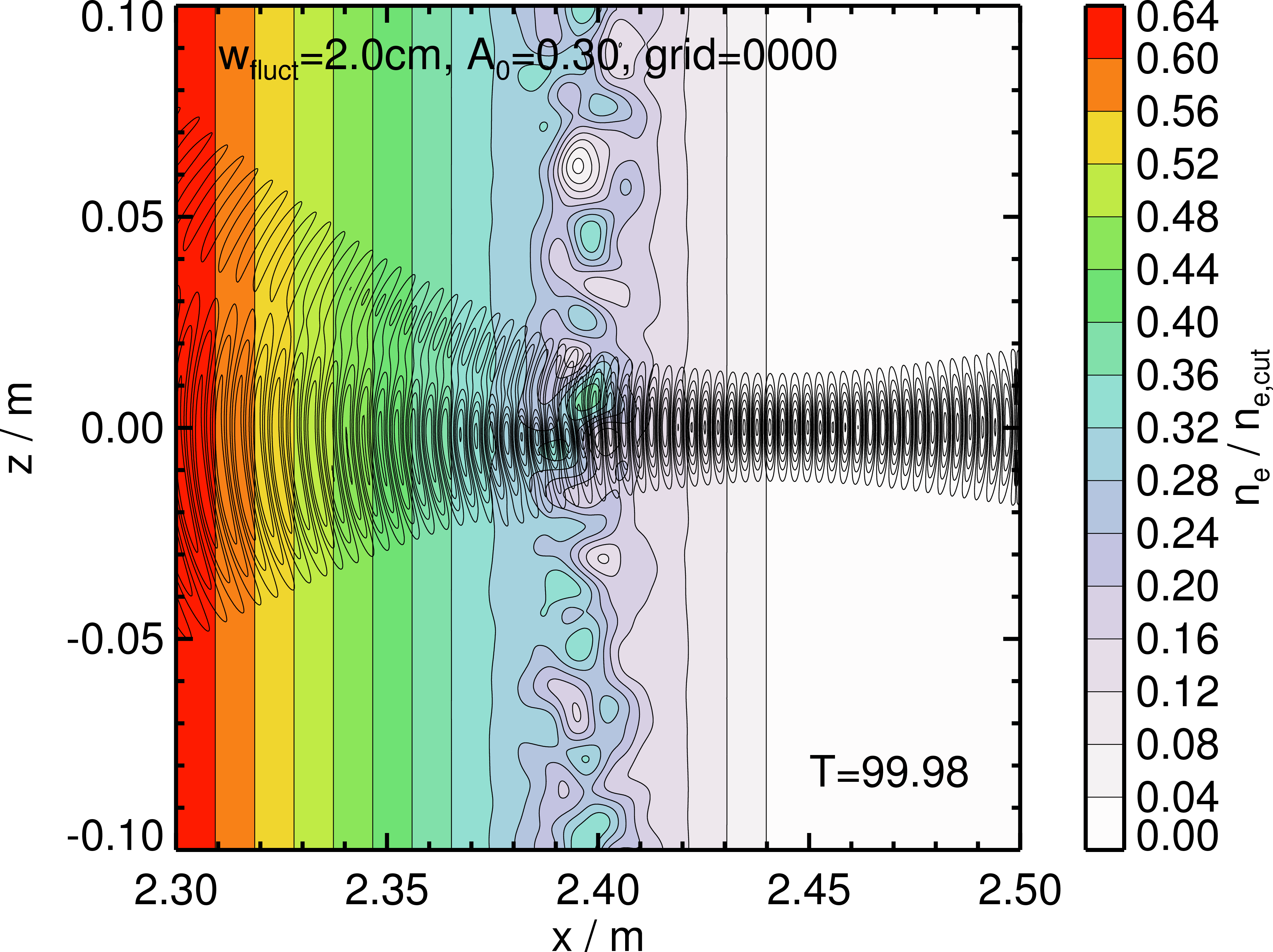}
\caption{Full-wave simulation grid with colored contour levels indicating the electron plasma density and contour lines indicating the absolute wave electric field. Shown is a sample from the ensemble with the average fluctuation parameters given in the plot. The snapshot is taken from a video published at~\cite{Koehn:2018b}.}
\label{f:movie_snapshot}
\end{figure}

The effect of a layer of turbulent plasma density fluctuations on a microwave beam can be nicely illustrated with full-wave simulations, as shown in Fig.~\ref{f:movie_snapshot}. The beam is injected from the right hand side and has its focal point approximately where the density profile starts to rise. A splitting of the injected beam is seen in this example, strongly perturbing the quality of the beam. The interested reader is referred to Ref.~\cite{Koehn:2018b} where the video belonging to the snapshot shown in Fig.~\ref{f:movie_snapshot} can be seen.

\begin{figure}[ht]
\centering
\includegraphics[width=0.45\textwidth]{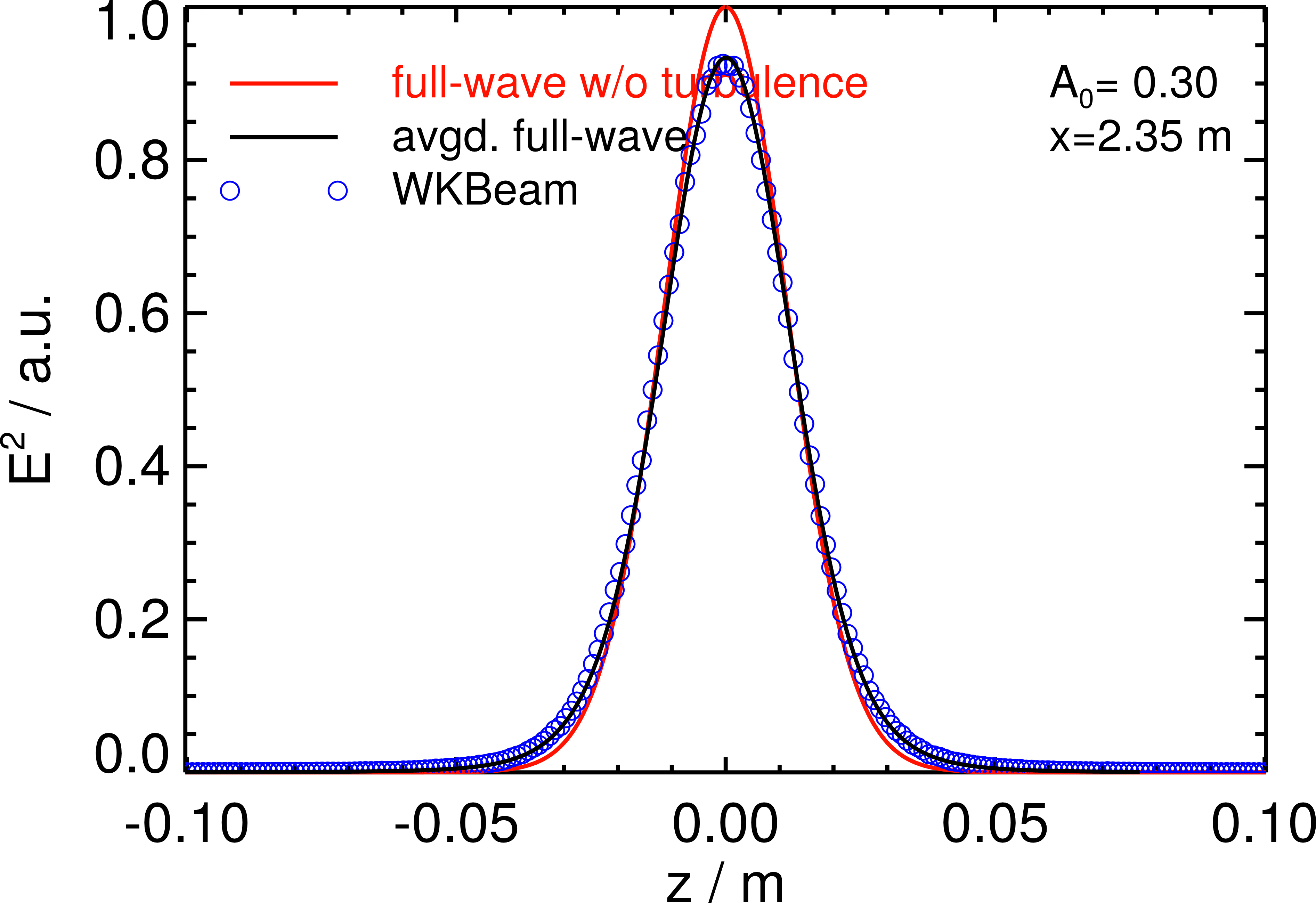}
\caption{Beam cross section at a radial position of $x=2.35\,\mathrm{m}$ from full-wave simulations (ensemble-averaged) and WKBeam calculations.}
\label{f:BenK_detAnt_Omode}
\end{figure}

For a quantitative data analysis, the average over the full ensemble of the full-wave simulations needs to be taken. This work uses an ensemble size of $N=3000$ which is large enough to ensure that the statistical error is small compared to the averages. Figure~\ref{f:BenK_detAnt_Omode} shows the ensemble-averaged beam cross section from full-wave simulations together with the corresponding result from WKBeam calculations. The turbulence parameters are the same as for the single sample shown in Fig.~\ref{f:movie_snapshot}. A first thing to notice is that the average beam does not change its direction of propagation. 
Hardly any difference of WKBeam to the full-wave solution can be seen in this representation. Both signals experience a small broadening (together with a reduction of the peak amplitude) as compared to the case without turbulence. 

\subsection{Scanning the fluctuation parameters}\label{s:BenK_scans}
In both codes, IPF-FDMC and WKBeam, the parameters defining the turbulent plasma density fluctuations were varied over a wide parameter range. 
This was done to explore the ability of the codes to analyze current and future devices, as well as to explore the validity of the Born approximation, expressed in Eq.~(\ref{e:wkbeam_born}), and thus of WKBeam.
The fluctuation strength, the background density, and the width of the fluctuation layer were varied in a series of parameter scans. Due to the limited space in this paper for the proceedings, we will only present a selection of the results here and like to refer the interested reader to the full paper~\cite{Koehn:2018a}.

\begin{figure*}[ht]
\centering
\includegraphics[width=0.95\textwidth]{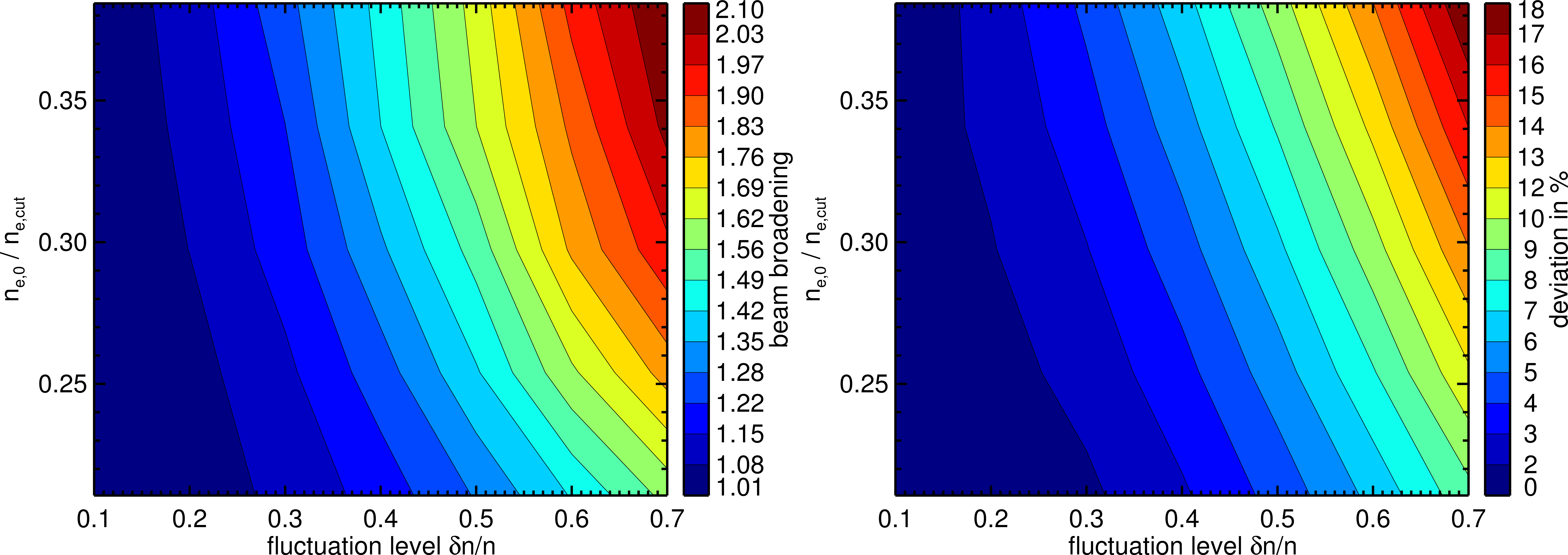}
\caption{(Left) Beam broadening as obtained from full-wave simulations as a function of the background density at the position of the fluctuation layer and as a function of the fluctuation level. (Right) Overestimation of the corresponding WKBeam results to the full-wave solution.}
\label{f:fullwave_ne0_nefluct_scan}       
\end{figure*}

Figure~\ref{f:fullwave_ne0_nefluct_scan} (left) shows the beam broadening deduced from full-wave simulations as a function of fluctuation level and background density at the location of the fluctuations. It can be clearly seen that the broadening is a function of both parameters, as expected. The contour plot on the right hand side shows the deviation of the corresponding WKBeam calculations with respect to the full-wave simulations. Note that WKBeam consistently overestimates the broadening. Even for large fluctuation levels of $50\,\%$, the disagreement to the full-wave solution is below $10\,\%$. Recalling the criterion to be fulfilled for the Born approximation to hold, Eq.~(\ref{e:wkbeam_born}), it is possible to draw a rough estimation: with the y-axis corresponding to the first factor and the x-axis to the second factor, the resulting product of the two quantities should be below $0.15$ to keep the overestimation of WKBeam below $10\,\%$.

Note that those results were obtained for a width of the fluctuation layer of approximately 
$w_0 = 2\,\mathrm{cm} \approx 3.3\,\lambda_0$, with $\lambda_0$ the vacuum wavelength of the injected microwave.
According to the results discussed in Ref.~\cite{Koehn:2016}, the beam broadening scales linearly with the width of the fluctuation layer if the fluctuation level is not too large. The observed scaling of the beam broadening may serve as estimations for current experiments or predictions for future experiments. 

An important result of the benchmark study is the applicability of WKBeam for modern tokamaks. For the ASDEX Upgrade tokamak, for example, edge density values reach at maximum $0.2\cdot n_{e,\mathrm{cut}}$~\cite{Willensdorfer:2012} and maximum fluctuation levels of $20\,\%$~\cite{Medvedeva:2017}. With the resulting control parameter given by Eq.~(\ref{e:wkbeam_born}) being below the stated value of $0.15$ (and by checking Fig.~\ref{f:fullwave_ne0_nefluct_scan} (right)) WKBeam can be reliably applied for this scenario. Considering ITER parameters~\cite{Snicker:2018}, the overestimation of WKBeam calculations is on the order of $1\,\%$ due to the correspondingly low (normalized) density. Note that the deviation of the WKBeam results does not depend on the width of the fluctuation layer~\cite{Koehn:2018a}. Due to the long propagation length of the injected microwave beam in ITER, this is an important information strengthening the reliability of WKBeam.

\subsection{Extrapolations towards ITER}
Mitigation and suppression of NTMs is of vital importance for future tokamak experiments as explained in the introduction. In ITER, the EC upper launcher system is designed for this purpose~\cite{Henderson:2008}. Recently, the importance of quantifying the influence of plasma density fluctuations on the efficiency of NTM stabilization in ITER has been highlighted~\cite{Poli:2015}. Due to the short spatial scales of the density fluctuations, their effect cannot be accounted for by tools based on the geometrical-optics approximations (which are otherwise commonly used to design and describe microwave heating scenarios in fusion experiments). The WKBeam code was deliberately developed to handle this. Estimating the beam broadening due to plasma density fluctuations in ITER using WKBeam is the topic of a paper published earlier this year~\cite{Snicker:2018}. The main result of that paper will only be briefly discussed here.

As shown in Section~\ref{s:BenK_scans}, the strength of the scattering of the microwave beam, precisely the beam broadening, depends on a number of parameters of the turbulent density fluctuations. Choosing a ''correct'' set of parameters is therefore important to get reliable results. The term ''correct'' refers to parameters describing the flat top phase of the ITER baseline scenario~\cite{Parail:2013}. The spatial profile of the fluctuations' amplitude and the perpendicular correlation length of the density structures used in the modeling are based on experimental surveys~\cite{Conway:2008}. The fluctuations amplitude profile we are using has a constant strength of $2\,\%$ in the core and $20\,\%$ in the scrape-off layer, corresponding to an H-mode discharge.

Comparing first the power deposition profiles deduced from WKBeam and from the beam tracing code TORBEAM for the case without fluctuations, they reveal excellent agreement. Including fluctuations, which cannot be done with codes like TORBEAM~\cite{Poli:2001}, a significant broadening of the deposition profile is found. Taking the effect of ballooning into account, i.e.\ a poloidal variation of the density fluctuation amplitude, an overall broadening of approximately a factor of 2 is obtained. Considering the installed microwave power in ITER, it can be concluded that the most dangerous NTMs expected to occur in the $15\,\mathrm{MA}$ $Q=10$ standard H-mode scenario can still be stabilized based on the current model of the density fluctuations. As mentioned in the beginning of this section, a detailed discussion of the results is found in Ref.~\cite{Snicker:2018}.

Besides the EC upper-launcher system, ITER is also equipped with an equatorial EC launcher~\cite{Henderson:2015} aiming to influence the sawtooth cycle which is thought to affect the onset of NTMs. It is thus of interest to estimate the deteriorating effect of density fluctuations on this scenario as well. The WKBeam code predicts a less severe beam broadening (referring to the upper-launcher calculations) of a factor of $1.2 - 2.5$, as discussed in detail in Ref.~\cite{Snicker:2018b}. Note that the broadening is also of less importance here as the requirement on the localization of the power deposition is reduced for influencing the sawtooth cycle.

\subsection{Cross-polarization scattering}
An interesting effect occurs at small plasma densities, where the dispersion surfaces of O- and X-mode are not well separated: mode coupling can occur. Triggered by fluctuations at low background plasma density values, this \emph{cross polarization scattering} can be problematic: mode scattered from one polarization to the other not only reduces the heating or current drive efficiency due to the original polarization but the other polarization can itself deposit power at non-optimal positions or locally damage wall components if being reflected at a cut-off. Thus, the scattering operator in WKBeam has been extended to also included the possibility of mode scattering~\cite{Snicker:2018b,Guidi:2016}. As the underlying assumptions for deriving the scattering operator are, however, only marginally fulfilled for the parameters where mode scattering can actually be relevant, a benchmark against full-wave simulations is required. Here, we present preliminary results of cross-polarization scattering obtained from full-wave simulations in the course of this (ongoing) benchmark study.

\begin{figure}[th]
\centering
\includegraphics[width=0.45\textwidth]{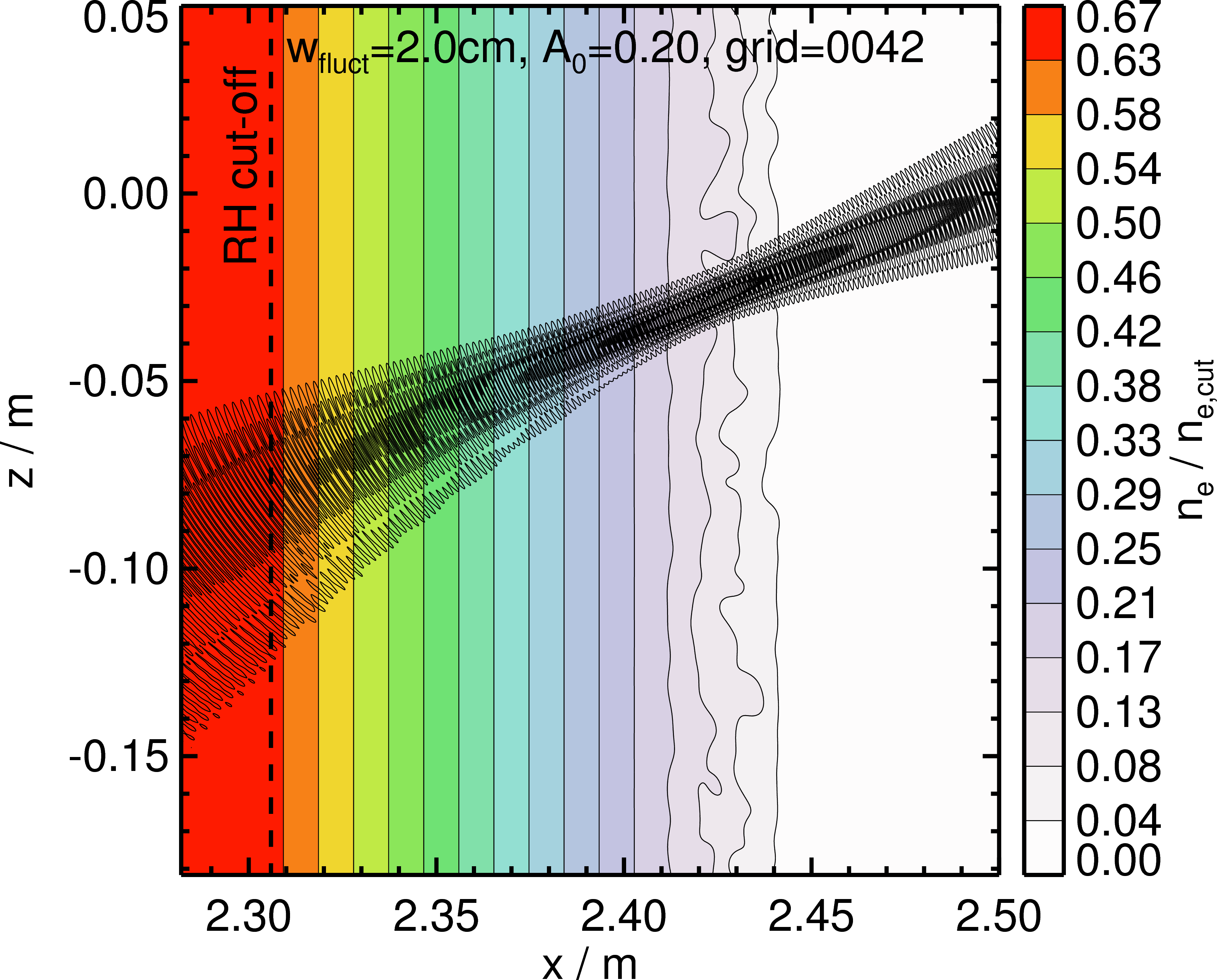}
\caption{Contour plot of the electron plasma density (colored) and of the absolute value of the wave electric field as obtained from full-wave simulations intending to study cross-polarization scattering due to density fluctuations at low plasma densities.}
\label{f:mode_scattering_example}
\end{figure}

The configuration of the simulations differs from the geometry described in Section~\ref{s:BenK_illustration}: it is optimized to result in a recognizable amount of mode-scattering. The wave frequency is increased to $110\,\mathrm{GHz}$ and the background magnetic field has a component pointing into the 2D simulation domain, $B_{\mathrm{tor}}=-1\,\mathrm{T}$, and a vertical component of $B_z=1\,\mathrm{T}$. The correlation length of the density structures is $L_\perp = 5\,\mathrm{mm}$ and the layer of fluctuations is shifted towards lower densities as shown in Fig.~\ref{f:mode_scattering_example}. Due to the oblique orientation of the background magnetic field, the injected O-mode is now elliptically polarized. As also shown in Fig.~\ref{f:mode_scattering_example}, the simulation domain includes the right-hand cut-off, which is a cut-off for the X-mode, in order to serve as a filter for that part of the microwave beam which is scattered from the injected O-mode to the X-mode: the X-mode is reflected at the cut-off and can thus be easily detected. To visually separate the reflected X-mode from the injected O-mode, an appropriate injection angle is chosen such that the two modes do not overlap.

Although no reflection at the right-hand cut-off is seen in Fig.~\ref{f:mode_scattering_example} at first glance, a small amount of cross-polarization scattering does actually occur in this case. It is approximately $0.40\,\%$ which is too small to be visible in this representation. As expected, even for the case without density fluctuations, there is a small amount of coupling to the other mode. From the full-wave simulations, a value of approximately $0.36\,\%$ is deduced for this geometry (which can be reduced by injecting wider beams with a narrower angular spectrum). The fluctuations thus result only in a small absolute increase of the wrong mode content. 

WKBeam calculations yield an X-mode content larger by approximately a factor $2-3$. Due to the assumptions being applied when deriving the scattering operator, this deviation is expected to change with the background density (into the direction that lower densities, when the modes are ''more'' degenerated, result in less reliable results). A systematic comparison between full-wave simulations and WKBeam is currently underway and will be published in a separate paper. 

As stated in Refs.~\cite{Snicker:2018b,Guidi:2016}, the cross-polarization scattering values obtained from WKBeam thus serve as an upper bound. Together with the preliminary results from full-wave simulations, cross-polarization due to density fluctuations seems to be on the same order in ITER as the polarization uncertainty due to the launcher. It is therefore not expected to result in severe problems. 

\section{Simulation-experiment comparison}
The microwave power deposition profiles can be experimentally determined using an electron cyclotron emission diagnostic in power-modulated discharges. An accompanying transport analysis separates the effect of profile broadening due to diffusion and due to other effects like a turbulent edge layer. Broadening of the injected EC beam of up to a factor of 2.5 could be determined in DIII-D this way, as presented on the last EC-conference~\cite{Brookman:2017}. Inspired by these experiments, we elaborated the possibility of performing simulations resembling the experiment as closely as possible. An ensemble of plasma density profiles including turbulence was created with the BOUT++ code~\cite{Dudson:2009} using the Hermes model~\cite{Dudson:2017}. Equilibrium profiles obtained from DIII-D discharges were used as input parameters. The fluctuation levels obtained from the BOUT++ simulations were confirmed to correspond to those found in the experiment. 

The ensemble of density profiles including plasma density fluctuations was then used as input for 3D full-wave simulations using the EMIT-3D code~\cite{Williams:2014}. The simulation domain does not include the full propagation path from the emitting antenna to the place of absorption in the plasma center. It is instead restricted to the area where the strongest fluctuation levels are observed as this region is thought to be responsible for the observed broadening. Making the assumption that the perturbed (ensemble-averaged) beam can be approximately described by a Gaussian (and that the simulation domain is sufficiently far away from the beam waist), its width can be linearly extrapolated to the location of absorption. Comparison with the experimentally deduced beam broadening is possible this way, yielding excellent quantitative agreement over a wide range of fluctuation levels (realized by different experimental scenarios). The results are presented and discussed in detail in a paper submitted recently~\cite{Brookman:2018}.

\section{Summary}
We have shown that turbulent edge density fluctuations can lead to significant broadening of an injected microwave beam. The stabilization of NTMs requires good spatial localization of the beam and the presented parameter scans of the fluctuation parameters may serve to quantitatively estimate the reduction in efficiency. Full-wave simulations have been used to benchmark WKBeam calculations, where very good agreement was found of a large parameter range. In particular for the expected density fluctuations in ITER, the deviations of the WKBeam are negligible. This supports the results from the corresponding calculations for ITER, performed with WKBeam, predicting that NTM stabilization is still possible in ITER taking into account the effect of fluctuations. A first estimation of the effect of cross-polarization scattering has been performed, which seems to not be an issue in ITER. Finally, a comparison between beam broadening obtained from a series of DIII-D experiments and from full-wave simulations of these experiments was presented yielding excellent agreement.

\section*{Acknowledgments}
This work has been carried out within the framework of the EUROfusion Consortium and has received funding from the Euratom research and training programme $2014-2018$ under grant agreement No.\ 633053. The views and opinions expressed herein do not necessarily reflect those of the European Commission.

The authors are indebted to the efforts of the open-source software community.


\end{document}